\documentclass[10pt,a4paper,notitlepage]{article}
\usepackage{amssymb}
\usepackage{amsmath}
\usepackage{amsfonts}
\usepackage{fancyhdr}
\usepackage{accents}
\usepackage{palatino}
\usepackage{setspace}
\renewcommand{\dot}[1]{\accentset{\mbox{\normalsize\bfseries .}}{#1}}
\renewcommand{\ddot}[1]{\accentset{\mbox{\normalsize\bfseries .\hspace{-0.10ex}.}}{#1}}

\begin{document}

\title{\textbf{Hyperspace fermions, M\"{o}bius transformations,\\ Krein space, fermion doubling, dark matter}}
\author{\textbf{George Jaroszkiewicz}\\
george.jaroszkiewicz@gmail.com}
\date{Final version: $7^{th}$ October 2020}
\maketitle

\begin{abstract}
We develop an approach to classical and quantum mechanics where continuous time is extended by an infinitesimal parameter $T$ and equations of motion converted into difference equations. These equations are solved and the physical limit  $T \rightarrow 0$   then taken. In principle this strategy should recover all standard solutions to the original continuous time differential equations. We find this is valid for bosonic variables whereas with fermions, additional solutions occur. For both bosons and fermions, the difference equations of motion can be related to M\"{o}bius transformations in projective geometry. Quantization via Schwinger's action principle recovers standard particle-antiparticle modes for bosons but in the case of fermions, Hilbert space has to be replaced by  Krein space. We discuss possible links with the fermion doubling problem and with dark matter. 
\end{abstract}

\section{Introduction}

This paper develops an approach to mechanics based on the following, referred to as the \emph{hyperreal strategy}:

\ 

\noindent \emph{``\ldots Hyperreals can also be used to find all the
solutions of the standard version of the Cauchy problem. First, perturb the
initial condition and/or the differential equation by an infinitesimal.
Then, find the unique solution to the hyperfinite difference equation using
the construction in the proof of PET (Peano's existence theorem). Finally,
take the standard part of the hyperreal solution.''} \hfill \hfill S. Wenmackers \cite%
{WENMACKERS-2012}

\ 

The differential calculus as developed independently by Newton and Leibniz
was not based on the rigorous $(\varepsilon ,\delta )$ arguments developed
later by Bolzano, Weierstrass and others \cite{GRABINER-1983}. The origins of
calculus were based on intuition and the heuristics of \emph{infinitesimals}%
, a philosophical and mathematical enigma dating from ancient times. The
success of $(\varepsilon ,\delta )$ in analysis, however (perhaps
unfortunately \cite{BLASZCZYK-2013}), gave rise to a widespread
view amongst  mathematicians that the concept of infinitesimal is ill-defined
and best avoided. A long standing question was: \emph{are infinitesimals no
more than a heuristic aid in certain calculations or can they be put on a
sound, rigorous footing}?

As in the case of the Dirac delta, the intuition behind infinitesimals was
eventually justified rigorously, with the development of nonstandard
analysis by Robinson \cite{ROBINSON-1966} and others, involving the
extension of the real number set $\mathbb{R}$ to the hyperreals, denoted $%
\mathbb{R}^{\ast }$.

Given that infinitesimals are mathematically legitimate, it seems reasonable
therefore to explore the strategy summarized by Wenmackers, above. The
differential equations we consider come from dynamical models related to
relativistic quantum field theory. The hyperreal strategy works precisely
for bosonic degrees of freedom. We will show, however, that in the case of
fermionic degrees of freedom, this approach leads to normal solutions plus a
bizarre kind of solution we encountered in our work on discrete time
mechanics \cite{B04_DTM}. These extra solutions, referred to here as \emph{%
hyperphase solutions}, do not have continuous time limits, existing in
states that, crudely speaking, oscillate in sign on infinitesimal scales.

A critical and essential point here is that hyperphase solutions, if and
when they occur, are legitimate mathematical solutions to hyperreal
difference equations. They cannot be dismissed simply because they have no
continuous time limits. Whether hyperphase solutions are relevant to physics
or not is, therefore, solely an empirical question. They do exist,
mathematically. Such solutions {might} model dark matter, if it could
be established that hyperphase solutions decouple from Maxwell fields and
cause curvature, which remains to be seen and is being investigated.

A significant feature of this approach is that we find hyperphase solutions
only in the case of fermionic degrees of freedom. The reason is that
differential equations of motion for fermionic fields are first order in
time whereas those for bosonic fields are generally second order in time. On
that account, we do not expect scalar, electromagnetic, non-abelian gauge
bosonic, or gravitational, field equations to support hyperphase solutions.

\section{Hyperreal numbers}

Hyperreal numbers are an extension of the real numbers that include
infinitesimals and infinitely large numbers \cite{KEISLER-2012,ROBINSON-1966}%
. These are `numbers' that satisfy all the standard rules of the reals plus
a few carefully chosen properties. For example, infinitesimals are hyperreal
numbers that satisfy the following condition: given any non zero
infinitesimal $T$ and any non zero real number $t$, then $0<|T|<|t|$. In
contrast, an infinitely large hyperreal is one that has a magnitude greater
than that of any real number. In our theory we shall make use also of
infinite hyperreal integers $n$. Such an integer $n$ satisfies the rule that
for any non-zero infinitesimal $T$, there exists a finite real $t$ such that 
$nT=t$. An important property of hyperreals that we use frequently is that
the product of an infinitesimal and a non-zero real is an infinitesimal.

In this paper we are concerned more with the application of hyperreals to
differential equations, rather than with their formal, specific mathematical
theory, which we assume is mathematically consistent in the way we apply it.
There are several variant approaches to infinitesimals and our usage is
based on the assumption that these approaches, though differing in certain
technical details, are all consistent with our usage.

Throughout this paper, the symbol $t$ will represent standard real,
continuous time. Given a non-zero, real infinitesimal $T$ and a
complex-valued function $f$ of $t$, we define a {hyperreal extension} $%
f_{T}$ of $f$ by the rule 
\begin{align}
f(t)\underset{T}{\rightarrow }f_{T}(t)\equiv & \ldots + \frac{f_{-1} (t)}{T}%
+f(t)+ f_{1}(t)T+f_{2}(t)T^{2}+\ldots ,
\end{align}%
where the coefficients $\ldots, f_{-1},f_1, \ldots$ are independent of $T$.
For those extensions with no negative powers of $T$, we define the {%
standard part} $\mathcal{S}_{T}f_{T}$ of $f_{T}$ with respect to $T$ by $%
\mathcal{S}_{T}f_{T}(t)=f(t)$.

If $f$ is a differentiable function of $t$, then we shall generally be
interested in hyperreal extensions of the form 
\begin{align}
f_{T}(t)\equiv f(t+T)=f(t)+\dot{f}(t)T+O(T^{2}),  \label{dfre}
\end{align}%
where $\dot{f}$ is the fluxion (conventional time derivative) of $f$.

The significance of taking standard parts is that according to the hyperreal
strategy, observable physics deals only with standard parts of hyperreal
extended equations and their solutions. This is analogous to working in
imaginary time (Euclidean field theory) and extracting physical predictions
in the real time limit.

In the following we use the symbol $\approx $ to express a Laurent or Taylor
series expansion in powers of $T$ up to some useful point, dropping the $%
O(T^{k})$ symbol (although it will always be implied). So for example
equation (\ref{dfre}) will be written $f_{T}(t)\approx f(t)$ when we wish to
ignore $O(T)$ terms and as $f_{T}(t)\approx f(t)+\dot{f}(t)T$ when we wish
to ignore $O(T^{2})$ terms.

In non-standard analysis (the mathematics of hyperreals), derivatives take
the form 
\begin{align}
\dot{f}(t)& \equiv \mathcal{S}_{T}\left\{ \frac{f(t+T)-f(t)}{T}\right\} ,
\qquad \ddot{f}(t) \equiv \mathcal{S}_{T}\left\{ \frac{f(t+T)-2f(t)+f(t-T)}{%
T^{2}}\right\} ,
\end{align}%
and so on. In our notation, we may write 
\begin{align}
f(t+T)-f(t) &\approx \dot{f}(t)T, \qquad f(t+T)-2f(t)+f(t-T) \approx \ddot{f}%
(t)T^{2},
\end{align}%
and so on.

In our approach, we hyperextend only in time and not in space. There are
several reasons for this but we will not review them at this point. 
What is important here is to investigate the {physical limit}, where it
exists, of hyperextended functions as $T$ is taken to zero with $t \equiv nT$
finite and fixed. For a hyperextended function $A(nT,T)$ we use the notation 
\begin{align}
\mathcal{P}_T A(nT,T) = \lim_{\substack{ T\rightarrow 0  \\ n\rightarrow
\infty  \\ nT=t \ \text{fixed}}}A(nT,T) \equiv A(t,0),
\end{align}
assuming $A(nT,T)$ is continuous in its second argument at $T=0$.

In our approach we encounter two kinds of temporal derivatives of
hyperextended functions. These are referred to as fluxions (standard time derivatives)
and hyperderivatives respectively, defined as follows. Given
a differentiable function $A(t, \boldsymbol{x})$ of normal spacetime
coordinates $(t,\boldsymbol{x})$, we will typically make a hyperreal
extension of the form 
\begin{align}
A(t,\boldsymbol{x}) \rightarrow A_n \equiv A(nT,T,\boldsymbol{x}),
\end{align}
where $n$ is an infinite hyperreal integer, $T$ is an infinitesimal, and $nT
= t$. Observable physics is defined by the physical limit of $A_n$ subject
to the constraint $nT = t$, where $t$ is a finite real number that plays the
role of continuous time. We define 
\begin{align}
A &\equiv \mathcal{P}_T \{ A_n \} , \qquad \dot{A} \equiv \mathcal{P}_T \{ 
\frac{\partial}{\partial t} A(t,T \}, \qquad A^{\prime} \equiv \mathcal{P}_T
\{ \frac{\partial}{\partial T} A(t,T \},
\end{align}
assuming these limits exist, and suppressing spacetime dependence in the
notation. There will be cases where such derivatives may not be always
exist, as occurs in the case of bosons.

From this, we deduce the rules 
\begin{align}
A_n & \approx A + {A}^{\prime} T, \qquad A_{n+1} \approx A + \dot{A} T +
A^{\prime} T , \qquad A_{n-1} \approx A -\dot{A}T +A^{\prime} T .
\end{align}

Such expansions involve the model parameters that specify and control
the dynamical equations concerned. In the case of dynamical variables%
, these by definition have only dynamical dependencies and no parametric
dependencies. So for a differentiable dynamical variable $q(t)$, bosonic or
fermionic, only fluxions occur and we have the relations 
\begin{align}
q_n & \approx q, \qquad q_{n+1} \approx q + \dot{q} T, \qquad q_{n-1}
\approx q -\dot{q} T .
\end{align}

\section{A first order example}

To illustrate the hyperreal strategy in operation and to understand what
happens with fermions, consider the first order ordinary differential
equation 
\begin{align}
\frac{d}{dt}f(t)=af(t),  \label{12345}
\end{align}%
where $f$ is a differentiable function of real time $t$ and $a$ is a finite
real constant. The standard solution is 
\begin{align}
f(t)=\mathrm{e}^{at}f(0).  \label{SOL1}
\end{align}

In the first instance, we would naturally assume that the hyperreal strategy
asserts that solutions to (\ref{12345}) should satisfy the equation 
\begin{align}
\mathcal{S}_{T}\left\{ \frac{f(t+T)-f(t)-Taf(t)}{T}\right\} =0,  \label{C}
\end{align}%
where $T$ is a non-zero infinitesimal.

Equation (\ref{C}) involves a {forwards difference quotient}, which is
a bias towards positive values of $t$. It is reasonable to question this
bias, because if $f(t)$ is a solution to (\ref{12345}) then we could equally
well assert that 
\begin{align}
\mathcal{S}_{T}\left\{ \frac{f(t)-f(t-T)-Ta f(t)}{T}\right\} =0,  \label{D}
\end{align}%
which involves a {backwards difference quotient}. We could even write 
\begin{align}
\mathcal{S}_{T}\left\{ \frac{f(t+T)-f(t-T)-2Ta f(t)}{2T}\right\} =0,
\label{E}
\end{align}%
which involves a {symmetric difference quotient}. Whilst none of the
statements (\ref{C}), (\ref{D}), and (\ref{E}) is problematical, the next
step in the hyperreal strategy requires further discussion. The strategy
requires us to remove the standard part operation and solve the resulting
difference equation. Removing the standard part operation gives the
following three difference equations respectively: 
\begin{eqnarray}
(\ref{C}) &\rightarrow &f_{n+1}=(1+aT)f_{n},  \label{C1} \\
(\ref{D}) &\rightarrow &f_{n+1}=(1-aT)^{-1}f_{n},  \label{D1} \\
(\ref{E}) &\rightarrow &f_{n+1}=2aTf_{n}+f_{n-1},  \label{E1}
\end{eqnarray}%
where $f_n \equiv f(t),f_{n+1}\equiv f(t+T), f_{n-1} \equiv f(t-T)$, and we
note that $aT$ is an infinitesimal.

The problem is this. Being first-order linear homogeneous difference
equations, (\ref{C1}) and (\ref{D1}) each have a unique solution for given
initial datum $f_0 \equiv f(0)$, these two solutions being different except
in the physical limit $T \rightarrow 0$. In contrast, equation (\ref{E1}) is
a {second} order difference equation, with {two} independent
solutions in general. The hyperreal strategy does not tell us explicitly how
to remove any second solution. To investigate this matter further, we now
solve each of these difference equations.

\ 

\noindent\textbf{Forwards differencing}: By inspection, equation (\ref{C1})
has solution$,$ 
\begin{align}
f_{n}=(1+Ta)^{n}f_{0}.  \label{AAA}
\end{align}%
Using the rule $\displaystyle{\lim_{n\rightarrow \infty }}\left(
1+x/n\right) ^{n}=\mathrm{e}^{x}$ then gives the required solution (\ref{SOL1}) in
the physical limit.

\ 

\noindent\textbf{Backwards differencing}: By inspection, equation (\ref{D1})
has solution 
\begin{align}
f_{n}=(1-aT)^{-n}f_{0},  \label{BBB}
\end{align}%
which also gives the required solution (\ref{SOL1}) in the physical limit.

It is important to observe that (\ref{AAA}) and (\ref{BBB}) are quite
different discrete functions but have the same physical limits. That
difference emphasizes the fact that discretization is not a unique process.
The issue we face is with the third possible hyperextension, (\ref{E1}).

\ 

\noindent\textbf{Symmetric differencing}: To solve (\ref{E1}), we assume a
solution of the form $f_{n}=z^{n}$. This gives the quadratic $z^{2}-2a T z-1=0$%
, which has standard solution $z=aT\pm \sqrt{1+a^{2}T^{2}}$. Hence there are
two independent solutions to our difference equation (\ref{E1}): 
\begin{align}
f_{n}^{(+)} =(aT+\sqrt{1+a^{2}T^{2}})^{n}f_{0}, \qquad f_{n}^{(-)} =(aT-%
\sqrt{1+a^{2}T^{2}})^{n}f_{0}.
\end{align}

Having found two independent solutions to our symmetric difference equation (%
\ref{E1}), we now apply the hyperreal strategy to recover the required
solution. It is at this point that the issue of two solutions to a first
order equation arises. Only the solution $f_{n}^{(+)}$ has a physical limit.
We find 
\begin{align}
\mathcal{P}_T f_{n}^{(+)} = \lim_{n\rightarrow \infty } \left\{1+\frac{at}{n}%
+O(\frac{1}{n^{2}})\right\}^{n}f_{0}=e^{at}f(0),
\end{align}%
recovering our required solution. However, 
\begin{align}
\mathcal{P}_Tf_{n}^{(-)} =\lim_{n\rightarrow \infty }\left\{ (-1)^{n}(1-\frac{at}{n}+O(%
\frac{1}{n^{2}}))^{n}\right\} f_{0},
\end{align}%
which does not exist. In heuristic terms, the solution $f_{n}^{(-)}$
oscillates in sign too rapidly as $n$ tends to infinity to have a physical
limit. Equivalently, we could say that the second solution changes sign on
infinitesimal scales and so should be unobservable by conventional means. We
shall refer to such rapidly oscillating solutions as \emph{hyperphase}
solutions.

Mathematically, we can rule out hyperphase solutions, if and when they
occur, by accepting only those solutions to our hyperreal difference
equations that have physical limits. However, mathematics is not physics, so
we should be prepared to give apparently spurious solutions to physical
equations some consideration as to whether they could indeed have a reasonable
physical interpretation. Such was the case, after all, with the negative
energy solutions to the Klein-Gordon equation, which led eventually to the
concepts of antiparticles and quantum field theory.

Having illustrated the hyperreal strategy, we turn now to practical
applications of it.

\section{The real bosonic oscillator}

In this section we apply the hyperreal strategy to the dynamical system
given by the Lagrangian 
\begin{align}
L=\tfrac{1}{2}m\dot{q}^{2}-\tfrac{1}{2}m\omega ^{2}q^{2},  \label{RTA}
\end{align}%
where the dynamical variable $q$ is real and bosonic, and where $m$ and $%
\omega $ are real constants.

We have found it necessary and unavoidable to deal with Lagrangians in the
first instance, rather than the equations of motion that they generate,
because physics involves more than just equations of motion. If we were
interested in coupling to gravitation, for instance, we would have to
discuss the energy-momentum stress tensor, as well as conserved
quantities such as electric charge. The best way of doing this is to first
hyperextend Lagrangians carefully, preserving whatever symmetries we
want to survive in the physical limit, and derive difference equations of
motion from those hyperextended Lagrangians. This naturally leads to the
technology of discrete time mechanics in the form discussed in \cite{B04_DTM}%
.

Given a continuous time Lagrangian $L$, the corresponding object in our
hyperreal discretization process is what we call a \emph{system function},
denoted $F_{n}$. System functions are, like Lagrangians, the keys to the
dynamics that they represent. Given a system function, we can construct
equations of motion and find invariants of the motion \cite{J1997D}. A
system function $F_{n}$ is a discrete time construct extending over the
temporal link $[nT,nT+T]$, satisfying the conditions 
\begin{align}
\mathcal{P}_{T}F_{n}\text{ } &=0,  \label{rrr} \\
\mathcal{P}_{T}\left( \frac{F_{n}}{T}\right) &=L,  \label{ddd}
\end{align}
where $L$ is the continuous time Lagrangian.
Before taking the physical limit, it is not necessary to think of $T$ as an
infinitesimal.

For the particular system of interest now, we deal with the variable $q_{n}$%
, where $n$ labels successive instants of time. These are separated by
intervals of time of duration $T$. Ultimately, $T$ will be taken to be an
infinitesimal, but in principle, could be a finite real number.

For the bosonic oscillator, experience leads us to define the system
function 
\begin{align}
F_{n}\equiv \tfrac{1}{2}A\left( q_{n}^{2}+q_{n+1}^{2}\right) -B q_{n}q_{n+1}
\label{Systemfunction}
\end{align}%
where $A $ and $B $ are real constants, with $B $ non-zero. Fixing the
hyperreal extension of these constants so as to lead to the continuous time
equations derived from (\ref{RTA}) is an important aspect of the hyperreal
strategy.

With a view to developments with the fermionic system discussed in a later
section, it is convenient to rewrite the above system function in the form 
\begin{align}
F_{n}=\tfrac{1}{2}B\boldsymbol{Q}_{n}^{\intercal}\mathbb{F}\boldsymbol{Q}_{n},
\end{align}%
where $\boldsymbol{Q}_{n}$  and $\mathbb{F}$ are given by 
\begin{align}
\mathbf{Q}_{n}\equiv 
\begin{bmatrix}
q_{n+1} \\ 
q_{n}%
\end{bmatrix}%
,\ \ \ \mathbb{F}\equiv 
\begin{bmatrix}
\eta & -1 \\ 
-1 & \eta%
\end{bmatrix}%
,
\end{align}%
$\boldsymbol{Q}_n ^{\intercal}$ is the transpose of $\boldsymbol{Q}_n$, and $\eta \equiv A/B $.

Equations of motion in this formalism are given by the
rule 
\begin{align}
\frac{\partial }{\partial q_{n}}\{F_{n}+F_{n-1}\}\underset{c}{=}0, \label{Cadzow}
\end{align}%
as discussed in \cite{B04_DTM}. Throughout this paper we use the symbol $\underset{c}{=}$ to denote {%
equality modulo equations of motion}. Applied to (\ref{Systemfunction}), rule (\ref{Cadzow}) gives
the equation of motion 
\begin{align}
q_{n+1}\underset{c}{=}2\eta q_{n}-q_{n-1},  \label{1111}
\end{align}%
which can be written in the form 
\begin{align}
\boldsymbol{Q}_{n}\underset{c}{=}\mathbb{E}\boldsymbol{Q}_{n-1},
\end{align}%
where $\mathbb{E}$ is the {matrix} 
\begin{align}
\mathbb{E}\equiv 
\begin{bmatrix}
2\eta & -1 \\ 
1 & 0%
\end{bmatrix}%
.
\end{align}

Anticipating the results of the {hyperreal extension} analysis
discussed below, we take $|\eta |\leqslant 1$ and write $\eta \equiv \cos
\theta $, where $\theta $ remains to be determined. Then $\mathbb{E}$ can be
written as 
\begin{align}
\mathbb{E\equiv }%
\begin{bmatrix}
\alpha +\beta & -\alpha \beta \\ 
1 & 0%
\end{bmatrix}%
,  \label{45678}
\end{align}%
where $\alpha $ and $\beta $ are the eigenvalues of $\mathbb{E}$, given by%
\begin{align}
\alpha \equiv \mathrm{e}^{\mathrm{i}\theta },\qquad \beta \equiv \mathrm{e}^{-\mathrm{i}\theta }.
\end{align}%
These eigenvalues are non-degenerate provided $\theta $ is not an integer
multiple of $\pi $, which has to be the case if we wish to recover our continuous time mechanics. We note that matrix $\mathbb{%
E}$ in the form (\ref{45678}) can be interpreted as a M\"{o}bius
transformation matrix in projective geometry, with fixed points $\alpha $
and $\beta $ and pole at zero. The same will be seen in our discussion of
fermions, below. The link between our hyperextension formalism and
projective geometry remains to be explored and should prove interesting.

The left-eigenvectors of $\mathbb{E}$ are 
\begin{align}
\boldsymbol{L}^{\alpha}\equiv 
\begin{bmatrix}
1, & -\beta%
\end{bmatrix}%
,\ \ \ \boldsymbol{L}^{\beta}\equiv 
\begin{bmatrix}
1, & -\alpha%
\end{bmatrix}%
.
\end{align}
Useful constructs corresponding to ladder (creation and annihilation)
operators in the quantized continuous time oscillator are given by 
\begin{align}
\mathcal{A}_{n}& \equiv \boldsymbol{L}^{\alpha }\boldsymbol{Q}%
_{n}=q_{n+1}-\beta q_{n,} \\
\mathcal{B}_{n}& \equiv \boldsymbol{L}^{\beta }\boldsymbol{Q}%
_{n}=q_{n+1}-\alpha q_{n,}.
\end{align}%
Then we find 
\begin{align}
\mathcal{A}_{n+1}\underset{c}{=}\alpha \mathcal{A}_{n},\ \ \ \mathcal{B}%
_{n+1}\underset{c}{=}\beta \mathcal{B}_{n},
\end{align}%
which can be used to solve the equations of motion completely.

A bilinear invariant readily found from the above, corresponding to the
conserved energy/Hamiltonian in the continuous time theory is given by 
\begin{align}
C_{n}\equiv \tfrac{1}{4}B\left( \mathcal{A}_{n}^{\intercal}\mathcal{B}_{n}+\mathcal{B}^{\intercal}%
_{n}\mathcal{A}_{n}\right) =\tfrac{1}{2}B\boldsymbol{Q}_{n}^{\intercal}{\mathbb{C}}%
\boldsymbol{Q}_{n},
\end{align}%
where $\mathbb{C}$ is the matrix 
\begin{align}
\mathbb{C}\equiv 
\begin{bmatrix}
1 & -\eta \\ 
-\eta & 1%
\end{bmatrix}%
.
\end{align}%
$C_{n}$ is conserved because we have the rule $\mathbb{E}^{\intercal}\mathbb{CE}=%
\mathbb{C}$.

The next step in the hyperreal strategy is to make hyperreal extensions of
all relevant quantities in our system function, as follows. Because we are
aiming to recover second order differential equations for $q(t)$, we make the
following hyperreal expansions: 
\begin{align}
q_{n}\approx q\equiv q(t),\qquad q_{n+1}\approx q+\dot{q}T+\tfrac{1}{2}\ddot{%
q}T^{2},\qquad q_{n-1}\approx q-\dot{q}T+\tfrac{1}{2}\ddot{q}T^{2}.
\end{align}%
With this and taking (\ref{ddd}) into account, we conclude that the
parameter $B$ requires a Laurent expansion in $T$ rather than a Taylor
series. Therefore, we write 
\begin{align}
B\approx \frac{B_{-1}}{T}+B_{0}+B_{1}T,
\end{align}%
where the coefficients $B_{i}$ are independent of $T$. By inspection, we
find we can get away with the parameter $\theta $ having a Taylor series
expansion, of the form 
\begin{align}
\theta \approx \theta _{0}+\theta _{1}T+\theta _{2}T^{2}.
\end{align}%
Then we find 
\begin{align}
F_{n}\approx \frac{\cos (\theta _{0})-1}{T}B_{-1}q^{2}+\ldots .
\end{align}%
Since we want (\ref{rrr}) to hold with both $B_{-1}$ and $q$ non-zero, we
set $\theta _{0}=0$. Then we find 
\begin{align}
F_{n}\approx \{\tfrac{1}{2}B_{-1}\dot{q}^{2}-\tfrac{1}{2}B_{-1}\theta
_{1}^{2}q^{2}\}T.
\end{align}%
Comparing this with (\ref{ddd}), we set 
\begin{align}
B_{-1}=m,\ \ \ \theta _{1}=\omega ,
\end{align}%
giving hyperreal consistency between our system function (\ref%
{Systemfunction}) and our original Lagrangian (\ref{RTA}).

The equation of motion (\ref{1111}) has hyperreal expansion 
\begin{align}
(m\ddot{q}+m\omega ^{2}q)T+O(T^{2})\underset{c}{=}0,
\end{align}
which is consistent with the harmonic oscillator equation derived from the
Lagrangian (\ref{RTA}).

For the ladder constructs, we find 
\begin{align}
\mathcal{A}_{n}\approx (\dot{q}+i\omega q)T,\qquad \mathcal{B}_{n}\approx (%
\dot{q}-i\omega q)T,
\end{align}%
consistent with the usual ladder operators. For the conserved quantity $C_{n}$ we find 
\begin{align}
\frac{C_{n}}{T}\approx \tfrac{1}{2}m\dot{q}^{2}+\tfrac{1}{2}m\omega
^{2}q^{2},
\end{align}%
which gives the correct energy in the physical limit.

Quantization can be done in two ways: canonical (operator) quantization or
Schwinger's source function approach. For bosons, canonical quantization is
straightforward and discussed next. We have found that Schwinger's approach
is best in the case of fermions, and we shall show how that works in the
next section.

For bosons, recall that in standard Hamilton-Jacobi theory, end-point
momenta are derived from Hamilton's principal function by the rule 
\begin{align}
p(t_{1})=-\frac{\partial }{\partial q(t_{1})}\int_{t_{1}}^{t_{2}}L dt,\ \ \
p(t_{2})=\frac{\partial }{\partial q(t_{2})}\int_{t_{1}}^{t_{2}}L dt.
\end{align}%
In discrete time mechanics, we have the analogous rule 
\begin{align}
p_{n}^{-}\equiv -\frac{\partial }{\partial q_{n}}F_{n},\ \ \ p_{n}^{+}\equiv 
\frac{\partial }{\partial q_{n}}F_{n-1}.
\end{align}%
This leads to the interpretation that the equation of motion (\ref{Cadzow})
expresses the equality of $p_{n}^{-}$ and $p_{n}^{+}$ over dynamical
trajectories. Our hyperreal expansions then give 
\begin{align}
p_{n}^{-}\approx m\dot{q},\qquad p_{n}^{+}\approx m\dot{q},
\end{align}%
as expected. Quantization of hyperextended variables is consistent with
standard canonical quantization if we adopt the rule 
\begin{align}
\lbrack p_{n}^{+},q_{n}]\underset{c}{=}[p_{n}^{-},q_{n}]\underset{c}{=}-\mathrm{i},
\end{align}
where we have set Planck's constant to unity for convenience.

It is a significant feature of our theory that bosonic variables are not
expected to have hyperphase modes. The reason is that bosonic particle and
field equations of motion in continuous time are second-order in general.
This includes standard gravitation. As we have seen, application of the hyperreal
strategy for bosons leads to second order difference equations, and these
will have two independent solutions in general. In the physical limit, one
solution corresponds to a positive energy solution propagating forwards in
time and the other corresponds to a negative energy solution propagating
backwards in time, corresponding to an antiparticle propagating forwards in time according to the Feynman-Stueckelberg interpretation of negative energy solutions.
Equivalently, these solutions correspond to standard quantum fields
propagating causally with Feynman propagators in the continuous time limit, rather than with retarded, advanced, or Dyson propagators. This is not the case for
fermions, as we shall show next.

\section{The fermionic particle}

\textbf{The continuous time model}

At this point we outline the continuous time model that we aim to
recover in our hyperextended formalism. There is one fermionic
(anticommuting) degree of freedom, $\psi (t)$ and its conjugate variable $%
\psi ^{\dagger }(t)$. Neither of these has any internal spin indices. The
Lagrangian, which incorporates a fermionic external source $\eta (t)$, is 
\begin{align}
L^{\eta }=\tfrac{1}{2}\mathrm{i}\psi ^{\dagger }\dot{\psi}-\tfrac{1}{2}\mathrm{i}\dot{\psi}%
^{\dagger }\psi -m\psi ^{\dagger }\psi +\eta ^{\dagger }\psi +\psi ^{\dagger
}\eta , \label{Lag}
\end{align}%
where $m$ is a mass. The equation of motion for $\psi $ is 
\begin{align}
\mathrm{i}\dot{\psi}-m\psi \underset{c}{=}-\eta . \label{sourced}
\end{align}

Applying Dirac's constraint approach to quantization \cite{DIRAC-1964} leads
to the quantum operator anticommutator 
\begin{align}
\{\psi ,\psi ^{\dagger }\}=1,  \label{444}
\end{align}%
where we have taken $\hbar =1$. Equivalently, applying Schwinger's action
principle 
\begin{align}
\delta \langle \Phi ,t_{2}|\Psi ,t_{1}\rangle _{\eta }\sim
\mathrm{i}\int_{t_{1}}^{t_{2}}dt\langle \Phi ,t_{2}|\delta L^{\eta}|\Psi ,t_{1}\rangle
_{\eta },\ \ \ t_{2}>t_{1}
\end{align}%
leads to the source-free vacuum expectation value%
\begin{align}
\langle 0_{+}|\mathcal{T}\psi ^{\dagger }(t_{1})\psi (t_{2})|0_{-}\rangle
= \mathrm{i}\Delta _{F}(t_{2}-t_{1})  \label{224}
\end{align}%
in Schwinger's notation \cite{SCHWINGER-1969}. Here $\mathcal{T}$ is the
usual time ordering operator, $|0_{-}\rangle $ and $|0_{+}\rangle $ are the 
\emph{in }and\emph{\ out }vacua, and $\Delta _{F}(t)$ is the Feynman
propagator for the system, given by%
\begin{align}
\Delta _{F}(t)=\mathrm{i}\mathrm{e}^{-\mathrm{i} mt}\theta (t).
\end{align}
Throughout this paper we assume $\langle 0_+ |0_- \rangle =1$. We can use (\ref{224}) to show that 
\begin{align}
\lim_{\varepsilon \rightarrow 0^{+}}\langle 0_{+}|\psi ^{\dagger
}(t+\varepsilon )\psi (t)|0_{-}\rangle &=0, \\
\lim_{\varepsilon \rightarrow 0^{+}}\langle 0_{+}|\psi (t)\psi ^{\dagger
}(t-\varepsilon )|0_{-}\rangle &=1,
\end{align}%
which is consistent with Dirac's operator quantization equation, (\ref{444}).

\

\noindent\textbf{The hyperparticle formalism}

We now apply the hyperreal strategy\ to the above continuous time model,
replacing temporal derivatives with appropriate hyperreal differences. We
define the discrete evolution operator $U_{n}$ and its inverse, $\overline{U}_n$, such that for any {%
normal} variable or function $O_{n}$ indexed by $n$, 
\begin{align}
U_{n}O_{n}\equiv O_{n+1},\ \ \ \overline{U}_{n}O_{n}\equiv O_{n-1}.
\end{align}%
Variables and functions that satisfy these relations will be referred to as 
\emph{normal}.

In our theory, not all variables turn out to be normal. Anticipating future
developments, we introduce the \emph{hyperphase symbol} $\xi $ with the
defining property that it commutes with everything except for the evolution
operators $U_{n}$ and $\overline{U}_{n}$. By definition, $\xi $ anticommutes
with those operators, so that for any normal indexed function or variable $%
O_{n}$, the product $\xi O_{n}$ is not normal. Specifically, we find 
\begin{align}
U_{n}(\xi O_{n}) &=-\xi U_{n}O_{n}=-\xi O_{n+1}, \\
\overline{U}_{n}(\xi O_{n}) &=-\xi \overline{U}_{n}O_{n}=-\xi O_{n-1}.
\end{align}%
Any object satisfying these last two conditions will be referred to as \emph{%
hypernormal.}

Our formalism was developed from the starting point that our variable $\psi
_{n}$ was normal. However, the solution $\psi _{n}\sim \beta ^{n}$ is
clearly hypernormal, because it does not have a physical limit. Taking the existence of normal and
hypernormal solutions into account we deduce that the variable $\Psi _{n}$
in our proposed equations has to be taken as a particular generalization,
that is, a combination of normal and hypernormal components. Therefore we
propose the expansions 
\begin{align}
\Psi _{n}& \equiv \psi _{n}+\xi \phi _{n}\approx \psi +\xi \phi \label{one}\\
\Psi _{n+1}& \equiv U_{n}\Psi _{n}=\psi _{n+1}-\xi \phi _{n+1}\approx \psi +%
\dot{\psi}T-\xi \phi -\xi \dot{\phi} T\label{two}\\
\Psi _{n-1}& \equiv \overline{U}_{n}\Psi _{n}=\psi _{n-1}-\xi \phi
_{n-1}\approx \psi -\dot{\psi}T-\xi \phi +\xi \dot{\phi}T. \label{three}
\end{align}%
In such an expansion, the components $\psi$ and $\phi$ are taken as normal, with good physical limits. Specifically, the component $\psi$ corresponds to the continuous time variable $\psi(t)$ occurring in Lagrangian (\ref{Lag}).  
We will refer to the variable $\Psi _{n}$ as a \emph{hyperparticle }in the
case of particle theories and as a \emph{hyperfield} when we are dealing
with fields (these are discussed in another article).

As in the bosonic case discussed in the previous section, we introduce a
fermionic bi-vector $\boldsymbol{\Psi }_{n}$ and its conjugate  $\boldsymbol{\Psi }_{n}^{\dagger}$ defined by 
\begin{align}
\boldsymbol{\Psi }_{n}\equiv 
\begin{bmatrix}
\Psi _{n+1} \\ 
\Psi _{n}%
\end{bmatrix}%
,\ \ \ \boldsymbol{\Psi }_{n}^{\dagger }\equiv 
\begin{bmatrix}
\Psi _{n+1}^{\dagger } & \Psi _{n}^{\dagger }%
\end{bmatrix}%
,
\end{align}%
where $\Psi _{n}$ is a hyperfermion.
With external fermionic sources defined by
\begin{align}
\boldsymbol{\eta}_{n}\equiv 
\begin{bmatrix}
\eta _{n+1} \\ 
\eta _{n}%
\end{bmatrix}%
,\ \ \ \boldsymbol{\eta}_{n}^{\dagger }\equiv 
\begin{bmatrix}
\eta _{n+1}^{\dagger } & \eta _{n}^{\dagger }%
\end{bmatrix},
\end{align}%
we consider the system function
\begin{align}
F_{n}=\boldsymbol{\Psi }_{n}^{\dagger }\mathbb{F}\boldsymbol{\Psi }_{n}+%
\tfrac{1}{2}T\boldsymbol{\eta}_{n}^{\dagger }\boldsymbol{\Psi }_{n}+\tfrac{1}{2}%
T\boldsymbol{\Psi }_{n}^{\dagger }\boldsymbol{\eta}_{n}.
\end{align}
Here $\mathbb{F}$ is the $2\times 2$ hermitian matrix 
\begin{align}
\mathbb{F}\equiv 
\begin{bmatrix}
A & -\mathrm{i}B^{\ast } \\ 
\mathrm{i}B & A%
\end{bmatrix}%
\end{align}%
where $B$ is complex and non-zero, $A$ is real, and $A$ and $B$ are constant in time $t$.

The hyperreal difference equations of motion are obtained by the same rule
as for the bosonic case discussed above, equation (\ref{Cadzow}),
giving the equation of motion 
\begin{align}
\Psi _{n+1}\underset{c}{=}2\mathrm{i} AB^{-1}\Psi _{n}+B^{-1}B^{\ast }\Psi
_{n-1}+\mathrm{i}B^{-1}T\eta _{n}. \label{FEQM}
\end{align}

Before we attempt quantization, we need to discuss the source free equation
of motion. In terms of the bivector notation, we write 
\begin{align}
\boldsymbol{\Psi }_{n}\underset{c}{=}\mathbb{D}\boldsymbol{\Psi }_{n-1},
\label{aaaa}
\end{align}%
where $\mathbb{D}$ is the matrix 
\begin{align}
\mathbb{D}\equiv 
\begin{bmatrix}
2\mathrm{i} AB^{-1} & B^{-1}B^{\ast } \\ 
1 & 0%
\end{bmatrix}%
.
\end{align}%
\qquad 

The two eigenvalues of $\mathbb{D}$ are 
\begin{align}
\alpha \equiv \frac{\sqrt{|B|^{2}-A^{2}}+\mathrm{i}A}{B},\ \ \ \beta \equiv -\frac{%
\sqrt{|B|^{2}-A^{2}}-\mathrm{i}A}{B}.
\end{align}

It is useful to reparametrize the parameters $A$ and $B$, noting that in the
physical limit we need $A^{2}<|B|^{2}$. Since $B$ is non-zero and
complex, we may write $B=|B|\mathrm{e}^{\mathrm{i}\delta }$ and $A=|B|\sin \theta $, where $%
\theta$ and $\delta$ are real. Then the eigenvalues take the form 
\begin{align}
\alpha =\mathrm{e}^{\mathrm{i}(\theta -\delta )},\ \ \ \beta =-\mathrm{e}^{-\mathrm{i}(\theta +\delta )}.
\end{align}%
A critical feature here is that these eigenvalues are not complex conjugates of each other. Examination of the physical limit shows that the region of interest in
this model is $A^{2}<|B|^{2}$, so we conclude that these eigenvalues are on
the unit circle and non-degenerate provided $\theta $ is not an odd multiple
of $\tfrac{1}{2} \pi$.
With this, we can write $\mathbb{D}$ in the form 
\begin{align}
\mathbb{D}\equiv 
\begin{bmatrix}
\alpha +\beta  & -\alpha \beta  \\ 
1 & 0%
\end{bmatrix}%
,
\end{align}%
which is in the form of a M\"{o}bius transformation, exactly as for the
bosonic model discussed above. The equation of motion (\ref{aaaa}) is then
equivalent to 
\begin{align}
\Psi _{n+1}\underset{c}{=}(\alpha +\beta )\Psi _{n}-\alpha \beta \Psi _{n-1}.
\label{457}
\end{align}%
As a second order difference equation, (\ref{457}) has two linearly
independent solutions, provided $\alpha \neq \beta $, which we assume. By
inspection, $\alpha \approx 1$, which means $\alpha ^n$ has a normal physical limit,  whilst $\beta^n $ is hypernormal because $\beta \approx -1$. This is analogous to the discussion of the first order differential  equation discussed in \S3 and is reflected in
the following analysis, particularly in our choice of propagator. In order
to recover the standard continuous time theory discussed above, we require
forwards in time propagation to be based on the $\alpha $ solution and not
the $\beta $ solution. This is analogous to working with the Feynman
propagator rather than the Dyson propagator.

Denote the $\alpha $-based solution by $\psi _{n}$ and the $\beta $-based
solution by $\xi \phi _{n}$, where we suppose that both $\psi _{n}$ and $%
\phi _{n}$ are differentiable functions of $t$ in the physical limit. Here $%
\xi $ is the hyperphase symbol introduced above.Then (\ref{457}) takes the
form 
\begin{align}
(U_{n}-\alpha -\beta +\alpha \beta \overline{U}_{n})\psi _{n}\underset{c}{=}%
\xi (U_{n}+\alpha +\beta +\alpha \beta \overline{U}_{n})\phi _{n}.
\end{align}%
Since the $\alpha $-based solution $\psi _{n}$ and the $\beta $-based
solution $\phi _{n}$ are supposed linearly independent, we require 
\begin{align}
(U_{n}-\alpha -\beta +\alpha \beta \overline{U}_{n})\psi _{n}& \underset{c}{=%
}0, \\
(U_{n}+\alpha +\beta +\alpha \beta \overline{U}_{n})\phi _{n}& \underset{c}{=%
}0,
\end{align}%
with the condition that both $\psi _{n}$ and $\phi _{n}$ have a normal
physical limit. The following analysis mirrors that for the bosonic system discussed above, at this point. 
The left-eigenvectors of $\mathbb{D}$ are 
\begin{align}
\boldsymbol{L}_{\alpha }\equiv 
\begin{bmatrix}
1 & -\beta 
\end{bmatrix}%
,\ \ \ \boldsymbol{L}_{\beta }\equiv 
\begin{bmatrix}
1 & -\alpha 
\end{bmatrix}.
\end{align}
The ladder variables are given by 
\begin{align}
\mathcal{A}_{n}& \equiv \boldsymbol{L}_{\alpha }\boldsymbol{\Psi }_{n}=\Psi
_{n+1}-\beta \Psi _{n}, \label{AA}\\
\mathcal{B}_{n}& \equiv \boldsymbol{L}_{\beta }\boldsymbol{\Psi }_{n}=\Psi
_{n+1}-\alpha \Psi _{n}. \label{BB}
\end{align}%
Significantly, these are not complex conjugates of each other because $\alpha$ and $\beta$ are not mutual complex conjugates.
The ladder variables satisfy the dynamical relations
\begin{align}
\mathcal{A}_{n+1}\underset{c}{=}\alpha \mathcal{A}_{n},\ \ \ \mathcal{B}%
_{n+1}\underset{c}{=}\beta \mathcal{B}_{n},
\end{align}%
which means
\begin{align}
\mathcal{A}_{n}=\alpha ^{n}\mathcal{A}_{0},\ \ \ \mathcal{B}_{n}=\beta ^{n}%
\mathcal{B}_{0}.
\end{align}%
From this, we can immediately write down two invariants of the motion:
\begin{align}
\mathcal{H}_{n}^{\alpha }\equiv \mathcal{A}_{n}^{\dagger }\mathcal{A}_{n}=%
\boldsymbol{\Psi }_{n}^{\dagger }\mathbb{H}^{\alpha }\boldsymbol{\Psi }_{n},%
\qquad \mathcal{H}_{n}^{\beta }\equiv \mathcal{B^{\dagger }}_{n}\mathcal{B}%
_{n}=\boldsymbol{\Psi }_{n}^{\dagger }\mathbb{H}^{\beta }\boldsymbol{\Psi }%
_{n},
\end{align}%
where
\begin{align}
\mathbb{H}^{\alpha }=%
\begin{bmatrix}
1 & -\beta \\ 
-\beta ^{\ast } & 1%
\end{bmatrix}%
,\ \ \ \mathbb{H}^{\beta }=%
\begin{bmatrix}
1 & -\alpha \\ 
-\alpha ^{\ast } & 1%
\end{bmatrix}%
.
\end{align}
We note the relations
\begin{align}
\mathbb{D}^{\dagger }\mathbb{H}^{\alpha }\mathbb{D=H}^{\alpha },\ \ \ 
\mathbb{D}^{\dagger }\mathbb{H}^{\beta }\mathbb{D=H}^{\beta }.
\end{align}

Although the operators $\mathcal{B}_{n}$ and $\mathcal{B}_{n}^{\dag }$ are
hyperphase operators, their bilinear combination $\mathcal{H}_{n}^{\beta }$
is normal. This suggests that hyperphase solutions could contribute to the
stress energy tensor, which is a source of gravitation. Further, if hyperphase matter
decoupled from the electromagnetic field but not to the stress-energy
tensor, then this could provide an explanation for dark matter.

We now consider quantization following Schwinger's functional approach
applied to discrete time mechanics. Equation (\ref{FEQM}) is now taken as an
operator equation of motion and written as
\begin{align}
\Psi _{n+1}\underset{c}{=}(\alpha +\beta )\Psi _{n}-\alpha \beta \Psi _{n-1}+%
\mathrm{i} TB^{-1}\eta _{n},
\end{align}%
where $\alpha $ and $\beta $ are as above. Schwinger's action principle in this context becomes
\begin{align}
\delta \langle \Phi ,N|\Psi ,M\rangle _{\eta }\sim
\mathrm{i}\sum\limits_{n=M}^{N-1}\langle \Phi ,N|\delta F_{n}|\Psi _{M}\rangle
_{\eta }.
\end{align}%
Following steps analogous to standard theory, we find%
\begin{align}
\mathcal{T}_{n,m}\langle 0_{+}|\Psi _{n}^{\dagger }\Psi _{m}|0_{-}\rangle
=-G_{m-n}, \label{timeordering}
\end{align}%
where $\mathcal{T}_{n,m}$ is the discrete time ordering operator and $G_{n}$
is a propagator satisfying the difference equation
\begin{align}
G_{n+1}-(\alpha +\beta )G_{n}+\alpha \beta G_{n-1}=B^{-1}\delta _{n}, \label{FPROP}
\end{align}%
with appropriate boundary conditions. In the following, $\theta _{n}$ and $%
\delta _{n}$ are discrete analogues of the Heaviside step $\theta (t)$ and
Dirac delta $\delta (t)$ defined as follows:%
\begin{align}
\theta _{n} &=\left\{ 
\begin{array}{l}
1,\ \ \ n=1,2,3, \ldots  \\ 
0,\ \ \ n<1%
\end{array}%
\right. , \qquad
\delta _{n} =\left\{ 
\begin{array}{c}
1,\ \ \ n=0 \\ 
0,\ \ \ n\neq 0%
\end{array}%
\right\} .
\end{align}%
Then taking the fermionic nature of the variables into account, we have
\begin{align}
\mathcal{T}_{n,m}\Psi _{n}^{\dagger }\Psi _{m}\equiv \Psi _{n}^{\dagger
}\Psi _{m}\theta _{n-m}+\tfrac{1}{2}\left\{ \Psi _{n}^{\dagger }\Psi
_{n}-\Psi _{n}\Psi _{n}^{\dagger }\right\} \delta _{n-m}-\Psi _{m}\Psi
_{n}^{\dagger }\theta _{m-n}.
\end{align}%
Because $\alpha^n$ behaves as a normal function having a proper physical limit, we impose the boundary condition that the $\alpha $ solutions propagate forwards in time, whereas the $\beta$ solutions  propagate backwards in time. Therefore, we
choose the conditions
\begin{align}
G_{n}\sim \alpha ^{n}, n\rightarrow +\infty ,\ \ \ G_{n}\sim \beta
^{n},n\rightarrow -\infty .
\end{align}%
Then the propagator $G_{n}$ satisfying (\ref{FPROP}) is given by
\begin{align}
G_{n}=\frac{1}{B(\alpha -\beta )}(\alpha ^{n}\theta _{n}+\delta _{n}+\beta
^{n}\theta _{-n}).
 \label{standardprop}
 \end{align}
Assuming that there is a physically meaningful Fock vacuum state, we define
the following vacuum expectation values:
\begin{align}
\langle 0_{+}|\Psi _{0}\Psi _{0}^{\dagger }|0_{-}\rangle & \equiv P,\ \ \ \
\ \langle 0_{+}|\Psi _{0}^{\dagger }\Psi _{0}|0_{-}\rangle \equiv Q, \\
\langle 0_{+}|\Psi _{1}\Psi _{0}^{\dagger }|0_{-}\rangle & \equiv R,\ \ \ \
\ \langle 0_{+}|\Psi _{0}\Psi _{1}^{\dagger }|0_{-}\rangle =R^{\ast }, \\
\langle 0_{+}|\Psi _{1}^{\dagger }\Psi _{0}|0_{-}\rangle & \equiv S^{\ast
},\ \ \ \ \ \langle 0_{+}|\Psi _{0}^{\dagger }\Psi _{1}|0_{-}\rangle =S.
\end{align}
where $P,Q,R$, and $S$ remain to be determined. Now if we were dealing with a standard Hilbert space, we would require both $P$ and $Q$ to be non-negative real numbers. We investigate this in the following three steps.

\

\noindent\textbf{1)} \ \
From the equations of motion
we find
\begin{align}
R& =(\alpha +\beta )P-\alpha \beta R^{\ast }, \qquad S =(\alpha +\beta )Q-\alpha \beta S^{\ast }.
\end{align}
\noindent\textbf{2)} \ \ Given the ladder operators defined by (\ref{AA}) and (\ref{BB}), we find
\begin{align}
\langle 0_+ |\mathcal{A}_{0}\mathcal{A}_{0}^{\dagger }|0_-\rangle & =2P-\beta R^{\ast
}-\beta ^{\ast }R, \qquad
\langle 0_+|\mathcal{A}_{0}^{\dagger }\mathcal{A}_{0}|0_-\rangle  =2Q-\beta S^{\ast
}-\beta ^{\ast }S,\\
\langle 0_+|\mathcal{B}_{0}\mathcal{B}_{0}^{\dagger }|0_-\rangle & =2P-\alpha
R^{\ast }-\alpha ^{\ast }R , \qquad
\langle 0_+|\mathcal{B}_{0}^{\dagger }\mathcal{B}_{0}|0_-\rangle  =2Q-\alpha
S^{\ast }-\alpha ^{\ast }S, \label{RHS}
\end{align}
Significantly, we find
\begin{align}
\langle 0_+|\mathcal{A}_{0}\mathcal{B}_{0}^{\dagger }|0_-\rangle & =
\langle 0_+|\mathcal{B}_{0}^{\dagger }\mathcal{A}_{0}|0_-\rangle  =
\langle 0_+|\mathcal{B}_{0}\mathcal{A}_{0}^{\dagger }|0_-\rangle  =
\langle 0_+|\mathcal{A}_{0}^{\dagger }\mathcal{B}_{0}|0_-\rangle = 0 \label{decouple}
\end{align}
The results of steps 1) and 2) are dependent only
on the equations of motion. Result (\ref{decouple}) means the $\mathcal{A}$ and $\mathcal{B}$ modes decouple and exist in disjoint sectors of their state space. Having such a decomposition is one of the defining properties of a Krein space \cite{BAGHERBOUM-2012}.

\

\noindent\textbf{3)} \ \ 
Finally, using (\ref{timeordering}) and (\ref{standardprop})
we find
\begin{align}
R &= \frac{\alpha }{B(\alpha -\beta )},\qquad
S =  \frac{\beta }{B(\beta -\alpha )},
\end{align}
and, contrary to expectations if we were dealing with a standard Hilbert space,
\begin{align}
P&=-Q=\frac{1}{B(\alpha -\beta )}.
\end{align}
With these relations we now find
\begin{align}
\langle 0_+|\mathcal{A}_{0}\mathcal{A}_{0}^{\dagger }|0_-\rangle & =\frac{(\beta -\alpha )}{B\alpha \beta }, \qquad
\langle 0_+ |\mathcal{A}_{0}^{\dagger }\mathcal{A}_{0}|0_-\rangle  =0,  \\
\langle 0_+ |\mathcal{B}_{0}^{\dagger }\mathcal{B}_{0}|0_-\rangle & =\frac{(\alpha -\beta )}{B\alpha \beta },\qquad \ 
\langle 0_+|\mathcal{B}_{0}\mathcal{B}_{0}^{\dagger }|0_-\rangle  =0.
\end{align}%
Hence we deduce%
\begin{align}
\langle 0_+ |\mathcal{A}_{0}\mathcal{A}_{0}^{\dagger }|0_-\rangle =-\langle 0_+ |\mathcal{B%
}_{0}^{\dagger }\mathcal{B}_{0}|0_-\rangle .
\end{align}

We interpret these results as follows.

\

\noindent\textbf{1.} \ \  The operator $\mathcal{A}^{\dagger}_n$ creates a normal fermion excitation propagating forwards in time, whereas $\mathcal{B}_n$ creates a hyperphase fermion excitation. 

\

\noindent\textbf{2.} \ \ The space $\mathcal{H}^{\alpha}$ of normal excitations has a positive definite inner product whereas the space $\mathcal{H}^{\beta}$ of hyperphase excitations has a negative definite inner product. Then the direct sum $\mathcal{H}\equiv \mathcal{H}^{\alpha} \oplus \mathcal{H}^{\beta}$ is a Krein space  \cite{BAGHERBOUM-2012}.

\

\noindent\textbf{3.} \ \ There is a notable history concerning indefinite metric quantum mechanics, with contributions from Dirac, Pauli, and many others. A recent discussion of the interpretation of such systems by Strumia, \cite{STRUMIA-2019}, focuses on the fact that an eigenstate of an operator, regardless of the metric, has a certain (that is, with probability of one)  outcome when acted on by that operator. According to Strumia, ``this is enough to make useful predictions even for non-trivial states''. It should be kept in mind that norms, inner products, indefinite metrics, quantum states, and so on, are all mathematical concepts. What matters is whether a given theory can predict empirically observable outcomes.  Few theorists would argue, for example, that the indefinite metric of Minkowski spacetime was devoid of physical significance.

\

To complete our analysis, we consider hyperextensions of all variables and parameters to find out what happens in the physical limit.
Assuming $\Psi _{n}$ can be
decomposed into two parts in the form given in  equation (\ref{one}),
where $\psi _{n}$ and $\phi _{n}$ have
physical limits, then with the hyperreal expansions (\ref{two}) and (\ref{three}),
and 
\begin{align}
\alpha & \approx 1+\alpha _{1}T,\ \ \ \beta \approx -1+\beta _{1}T,
\end{align}
where $\alpha_1$ and $\beta_1$ are to be determined, 
we find
\begin{align}
\dot{\psi}\underset{c}{=}\alpha _{1}\psi ,\ \ \ \dot{\phi}\underset{c}{=}%
-\beta _{1}\phi .
\end{align}
Taking $\alpha _{1}=-\mathrm{i}m$ recovers the normal free particle solution
\begin{align}
\psi (t)=\mathrm{e}^{-\mathrm{i} mt}\Psi (0).
\end{align}%
If we take $\beta _{1}=\mathrm{i}\mu $, then the hyperphase field $\phi _{n}$
propagates (in hyperspace) as a \textquotedblleft normal\textquotedblright\
field of mass $\mu $, where $\mu $ could be chosen different to $m$. That $\mu$ need not be equal to $m$ tells us that hyperphase solutions do not correspond to antiparticles as they are understood conventionally. 

\

Note that we need to take  $B\approx \tfrac{1}{2}$ in order to recover the
sourced equation of motion (\ref{sourced}).

\

A final point is that in the physical limit, we find  $\mathcal{A}_{n}\approx 2\psi$ and
\begin{align}
\langle 0_+ | \mathcal{A}_{n}\mathcal{A}_{n}^{\dagger }|0_-\rangle & \approx 4,
\end{align}
consistent with the original continuous time model discussed at the start of this section.

\section{Concluding remarks}
There are several points that should be commented on concerning the hyperreal approach to mechanics.

\

\noindent\textbf{Mathematical consistency} 

\noindent This paper explores the tension between two different mathematical ideologies, namely, the $(\varepsilon ,\delta)$ approach to calculus versus the infinitesimal approach to calculus. This tension leads naturally to an approach to mechanics that fuses purely mathematical issues and purely physical issues.   On the mathematical side, we have the hyperreal strategy approach to differential equations, as succinctly expressed by Wenmackers and quoted at the start of this paper. On the physics side, we found that we had to apply that strategy at the Lagrangian level rather than to equations of motion directly. That fusion gives an inevitability to the hyperphase modes we are reporting here. Before the physical limit is taken, hyperphase solutions exist as legitimate mathematical solutions to fermionic difference equations of motion and cannot be dismissed by fiat.  Given that the normal modes are physical, this then raises the question of whether the hyperphase fermionic modes could have physical significance. 

\

\noindent\textbf{Fermion doubling} 

\noindent There is an interesting parallel and possible link  here with the problem of fermion doubling
in lattice gauge theory. In that approach to hadronic physics, Lagrangians for chiral fermions are rewritten over finite, discretized spacetime lattices.
The imposition of periodic boundary conditions then appears to lead to spurious
solutions, creating the notorious fermion doubling problem. The reasons for this have been attributed to a combination of non-uniqueness in the discretization procedure, periodic boundary conditions, and chirality. In our approach,
we find that hyperphase solutions for fermions occur regardless of
any spinorial or chiral properties of the fermions, and we do not impose
periodic boundary conditions. 

It might be argued, as it is argued in the case of lattice field theories, that hyperphase solutions are artefacts of the particular
discretization process employed. Our view is that although discretization is not a unique process, the common factor responsible for the phenomenon we are reporting and the fermion doubling problem is that in each case, it is Lagrangians that are being discretized and not just equations of motion. Lagrangians are involved in lattice gauge theories because path integrals (more correctly, path summations) are used to calculate amplitudes, and that involves Lagrangians. In particular, Lagrangians are bilinear in fermionic variables, and that is the root of the problem as far as our approach is concerned. We do not need to discuss spin, chirality, or periodic boundary conditions to encounter hyperphase solutions.  

\

\noindent\textbf{Zitterbewegung} 

\noindent One referee asked about the relation between our hyperphase solutions and the phenomenon of Zitterbewegung, the rapid oscillatory motion of particles described via relativistic wave equations. Both phenomena involve rapid sign changes, but the similarities end there. Zitterbewegung can be thought of as an incoherent fluctuation of particle position, whereas hyperphase solutions exhibit coherent sign changes. Zitterbewegung is associated with a finite characteristic frequency of the order $m c^2 / \hbar$, whereas in the physical limit, hyperphase solutions occur on an infinitesimal time scale, which is by definition, not measurable.  

\

\noindent\textbf{Ghosts and negative inner products} 

\noindent At first sight, quantized hyperphase modes have some of the characteristics of the ghost fields  encountered in several contexts in quantum field theory. Perhaps the best known of these contexts is the Gupta-Bleuler approach to the quantization of Maxwell fields \cite{GUPTA-1950,BLEULER-1950}. There, time-like polarization modes have negative norms and, crudely speaking, cancel spatial longitudinal modes, giving 
physically observable photon states with two transverse polarization degrees of freedom. In the present case, hyperphase modes appear even though our fermionic degrees of freedom carry no spin polarization degrees of freedom. 

Ghost fields are also encountered in various approaches to quantum field theoretic path integrals, where they are used to encode gauge constraints. Whilst the hyperreal approach to gauge field theory remains to be explored, the hyperphase modes encountered in the model discussed above appear to have nothing to do with gauge theory.

\ 

\noindent\textbf{Potential application to dark matter physics}

\noindent In recent years it has become abundantly clear from astrophysical evidence that there is a strange form of matter permeating the universe that couples to gravitation but not to electromagnetism directly. At present, there is intense speculation about what dark matter is.  It has occurred to us  that dark matter might be explained as hyperphase fermionic matter. According to our thinking, hyperfields occurring in system functions and Lagrangians occur bilinearly. Given that normal and hypernormal modes decouple, then the hyperphase symbol $\xi$ should not appear in whatever energy-momentum-stress tensor was being discussed as a source of gravitation. Therefore, hypernormal modes could contribute to gravitational curvature, as well as normal modes. It remains to be seen whether such fields could be arranged to decouple from Maxwell fields in the negative inner product sector of the Krein space involved. What encourages us in this speculation is that we have seen above that the mass $\mu$ of the hyperphase component field $\phi$ need not be the same as the mass $m$ of the normal field $\psi$.

We do not wish to overstate the possibilities. As pointed out by one referee, it is fairly easy to propose novel particle fields as an explanation of dark matter, but working out the specific details that fit the currently available empirical data is a much more complex task. We would emphasize the point, however, that the approach outlined in this paper really does not fit into the category of an ad-hoc model deliberately designed to fit dark matter data. Rather, hyperphase solutions are unavoidable mathematical consequences of the hyperreal strategy applied to fermions, and were encountered before any connection with dark matter was contemplated.  Having found those solutions and examined their properties, it then seems reasonable to investigate whether they could indeed ``explain'' dark matter.

Work is in hand applying the hyperreal strategy to non-abelian fermion quantum field theory in 1+3 spacetime dimensions, in order to ascertain the possibilities in that respect.     
There are two empirical facts that appear consistent with, and indeed helpful to, the application of the hyperreal strategy to 1+3 dimensional spacetime physics.  First there is the question of which local inertial frame in which to apply the strategy.  Empirical physics helps us here, because observation of the cosmic background microwave radiation field shows that there is a local, preferred frame of rest, the local symmetry rest frame (modulo spatial rotations) of that cosmic background radiation field. The observed anisotropy in that radiation field, known as the dipole effect,  is generally interpreted  as a signal that our Galaxy is moving at about 600 km/sec with respect to that frame in the direction of the constellation Leo. In the first instance, we would apply our hyperreal strategy in that local symmetry rest frame.

Second, the Minkowski metric specifically singles out one of the four spacetime dimensions in any given inertial frame. Therefore, applying the hyperreal strategy to that dimension and not to the other three seems reasonable. Time is, after all, not space.  

Preliminary calculations show that the hyperreal strategy applied to the Dirac field in 1+3 dimensions recovers standard Lorentz covariant quantum field theory in the physical limit, plus hypernormal solutions analogous to the hypernormal solutions found in the present paper.

\end{document}